\documentclass[fleqn,12pt,twoside]{article}
\usepackage{espcrc1}

\usepackage{graphicx}
\usepackage[figuresright]{rotating}

\newcommand{\AmS}{{\protect\the\textfont2
  A\kern-.1667em\lower.5ex\hbox{M}\kern-.125emS}}

\newcommand{\eqn}[1]{\label{eq:#1}}

\def\veft{$V_{EFT}$}
\def\vph{$V_{ph}$}

\def\1s0{{}^1S_0}
\def\ts1{{}^3S_1}
\def\td1{{}^3D_1}
\def\ltap{\ \raise.3ex\hbox{$<$\kern-.75em\lower1ex\hbox{$\sim$}}\ }
\def\gtap{\ \raise.3ex\hbox{$>$\kern-.75em\lower1ex\hbox{$\sim$}}\ }
\def\ket#1{\vert#1\rangle}
\def\bra#1{\langle#1\vert}

\title{Renormalization group analysis of nuclear force}

\author{
S.X. Nakamura\address[triumf]{
Theory Group, TRIUMF, 
4004 Wesbrook Mall, Vancouver, BC V6T 2A3, Canada}}

\begin{document}

% typeset front matter
\maketitle

\begin{abstract}
We study a relation between nuclear forces based on phenomenological
approach (\vph)
and nuclear effective field theory (\veft) from a viewpoint of renormalization
group. We find
the relation between these two types of nuclear force using Wilsonian
renormalization group equation. Considering the fact that \veft\ is defined
in a certain
small model space, we show that a simple contact interaction accurately
simulates
short-distance physics, and that \veft\ is free from dependence on
 modelling the details of the short-distance physics.
Based on our result, we discuss some features of nuclear effective field
 theory.
\end{abstract}

\section{Introduction}
Nuclear force is one of the oldest problem in nuclear physics. However,
its theoretical
description is still an unsettled issue. Here, we first explain a traditional,
phenomenological approach to the nuclear force, and a fairly new approach
based on nuclear effective field theory (NEFT)\cite{weinberg}. After the explanation, we formulate a
scenario for a
relation between nuclear forces based on the two approaches. The main
subject in this report is to discuss the relation from a viewpoint of
the renormalization group (RG).

In the phenomenological nuclear force (\vph),
the long-range part is described by the
one-pion-exchange potential (OPEP). The short-range part is poorly known, and
therefore a
phenomenological model is used; heavy-meson exchanges or purely
phenomenological
parameterization. Parameters involved in the model are fixed so as to
reproduce
low-energy $NN$ scattering data and the deuteron binding energy. In this 
way,
several high-precision $NN$ potentials have been constructed, such as
the CD-Bonn and the Nijmegen
potentials. Although these models are phenomenologically very
successful, there are some problems as follows.
At first, each model describes the short-range part
differently, and is
largely model-dependent. Secondly, there is no systematic way
to construct
the short-range mechanism. Thirdly, there is little connection to the
underlying theory, QCD.

In contrast,
it is claimed that a nuclear force based on NEFT (\veft) is free from
the problems inherent in \vph. 
The claim is based on the derivation procedure of
\veft, which is briefly given as follows.
One starts with an effective chiral Lagrangian, taking care of 
the spontaneously broken chiral symmetry of QCD.
The Lagrangian consists of effective degrees of freedom for a system in
question, and is the most general as long as
assumed symmetries are satisfied.
One identifies a set of irreducible diagrams from the Lagrangian with 
\veft. The importance of an irreducible diagram is assigned by a
counting rule.

However, one may find questions about \veft:
(Q1) In \veft, short-distance physics is described by contact
interactions, which is much simpler compared to the short-range
mechanism employed in \vph. Is the contact interaction really
appropriate to describe the short-distance physics?
(Q2) Is \veft\ still one of many phase-equivalent potentials, as
\vph\ is?
(Q3) Is there any relation between model independent \veft\ and model
dependent \vph?
In fact, the answer to (Q3) naturally leads us to the answers to (Q1)
and (Q2). 
Thus, we formulate a scenario for the relation between \veft\ and \vph.

It is noted that the model space (the state space for the nucleon) of
\veft\ is considerably smaller than  
that of \vph. With this point in mind, we can formulate the following scenario
for the relation between \veft\ and \vph.
We may construct many \vph\ which reproduce $NN$ scattering data and the
deuteron binding energy.
They are different in describing the short-distance physics, and therefore
model dependent. 
Starting with such \vph, we reduce their model space by integrating out
the high momentum states of the nucleon.
Reducing the model space corresponds to viewing the system in a 
coarse-grained manner. As the model space is reduced, information about
details of the short-distance
physics is gradually lost. Eventually, we obtain a low-momentum effective
interaction ($V_M$) defined in a reduced model space. $V_M$ does not have
the model dependence which \vph\ have. 
A parameterization of $V_M$ constitutes \veft.
This is the scenario for the relation between \veft\ and \vph.
\veft\ obtained in this way is, by construction, does not have a
dependence on modelling the short-distance physics.
The short-range part of $V_M$ is expected to be accurately simulated by
simple contact interactions because the detailed information has been
integrated out.
Therefore, if we show that this scenario is realized, then we can answer
all of the questions raised in the previous paragraph.

The purpose of this report is to confirm the scenario, thereby proposing 
the relation between \veft\ and \vph\cite{mine}.
Furthermore, keeping the relation in mind, we can
understand features of NEFT more deeply, which we will discuss later.

\section{Wilsonian renormalization group equation}
In order to examine the scenario, stated in the previous section, for the
relation between
\veft\ and \vph, an appropriate model-space reduction method is necessary.
In effective filed theory, we use an effective Lagrangian in which
high-energy degrees of
freedom have been integrated out using a path integral:
\begin{eqnarray}
\eqn{z_ll}
 Z &=& \int {\cal D}\Psi_H{\cal D}\Psi_L\,
 e^{i\int  d^4\!x\ {\cal L}_H}= \int {\cal D}\Psi_L\, e^{i\int  d^4\!x\
 {\cal L}_L}\ ,
\end{eqnarray}
where $\Psi_H$ ($\Psi_L$) is high-(low-) energy degrees of freedom, and
${\cal L}_H$ (${\cal L}_L$) consists of $\Psi_H$ and $\Psi_L$ (only
$\Psi_L$). An effective (more fundamental) Lagrangian is ${\cal L}_L$
(${\cal L}_H$).
The model-space reduction, integrating out high-momentum states of the
nucleon, also should be
done with the path integral. Although it is difficult in general, in case
the Lagrangian
is composed by only the nucleon field and two nucleons interact through
contact
interactions, we can perform the path integral. 
It is noted that we restrict ourselves to two-nucleon system throughout
this report. In the center of mass
frame, performing the path
integral is equivalent to solving the following Wilsonian renormalization
group (WRG) equation:
\begin{eqnarray}
\eqn{rge}
 {\partial V^{(\alpha)}(k',k;p,\Lambda) \over\partial\Lambda}
= {M\over 2\pi^2} V^{(\alpha)}(k',\Lambda;p,\Lambda){\Lambda^2\over\Lambda^2-p^2}
V^{(\alpha)}(\Lambda,k;p,\Lambda)\ ,
\end{eqnarray}
where $V^{(\alpha)}$ is $NN$ potential for a partial wave $\alpha$,
and $M$ the nucleon mass.
In $V^{(\alpha)}(k',k;p,\Lambda)$,
$k$ ($k'$) denotes the off-shell 
relative momentum of the two nucleons before
(after) the interaction, $p$ denotes the on-shell momentum, and
$\Lambda$ is the cutoff value specifying the model space.
This equation controls the evolution of $V^{(\alpha)}$ with respect to a
change of $\Lambda$.
This equation was firstly derived by Birse {\it et al.} in a different
manner\cite{birse}.

\section{Result}

Starting with several phenomenological $NN$ potentials, we reduce
their model spaces using the WRG equation, and examine the evolution.
Then, we simulate the obtained effective model-space
interaction ($V_M$) with \veft\ by adjusting parameters in it.
If the simulation is accurate, then the relation between \veft\ and
\vph\ through RG is realized.

In the simulation, for simplicity, we include only the OPEP in \veft\ as
a mechanism explicitly including the pion. 
This is not fully
consistent with Weinberg's counting.
Because we consider contact interactions with zero, two, and
four derivatives,
we should include more irreducible graphs,
such as a two-pion exchange potential (TPEP).
However, 
we employ a rather small model space where the
details of the TPEP play essentially no role; the TPEP is accurately
simulated by contact interactions.
Therefore, our simplification does not deteriorate the accuracy of the
simulation.

\begin{figure}[h]
 \begin{minipage}[t]{75mm}
 \includegraphics[width=75mm]{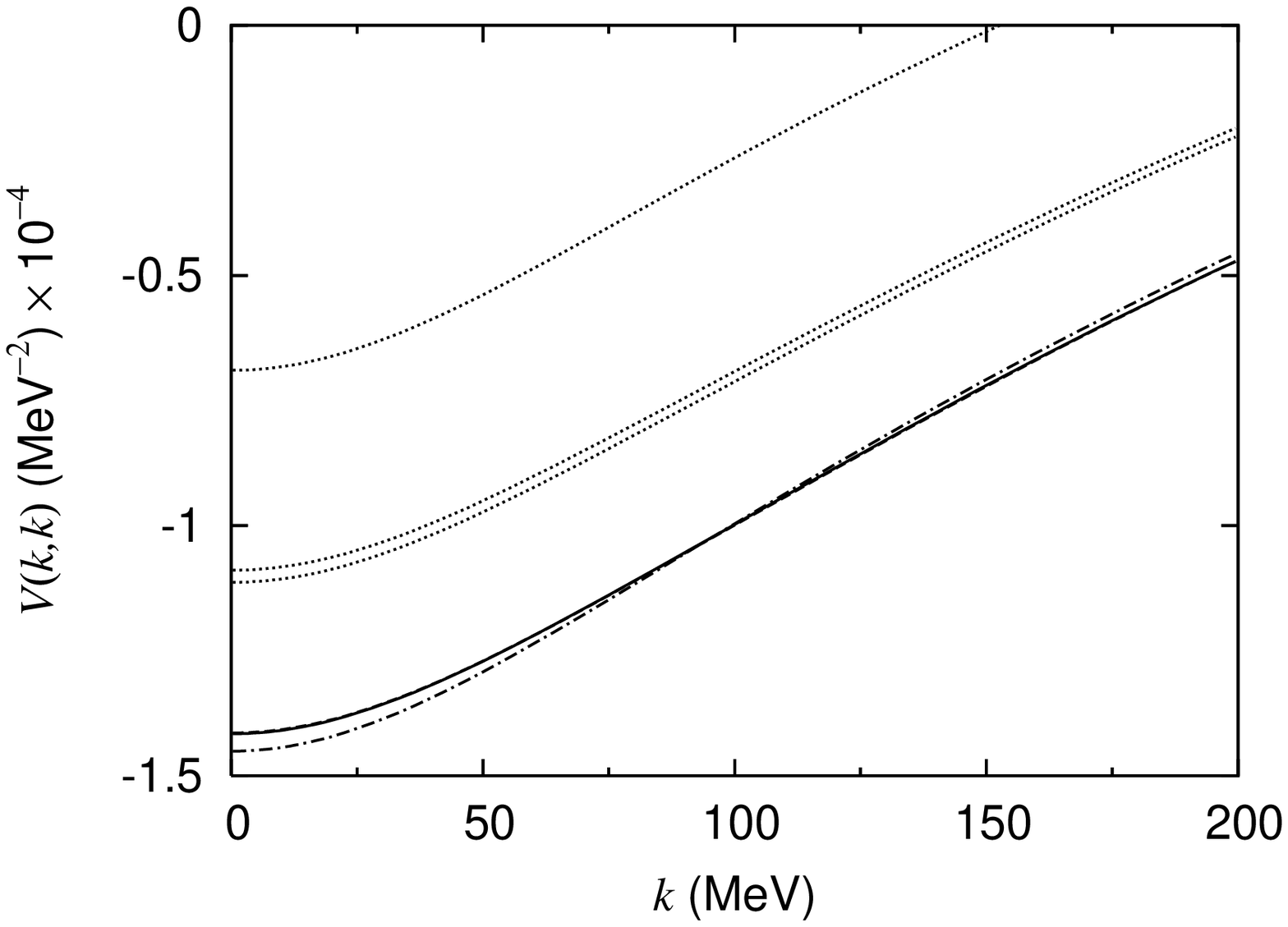}% Here is how to import EPS art
 \caption{\label{fig1}
$NN$-potential
$\bra{\1s0}V\ket{\1s0}$;
the upper,
the Reid93, the Nijmegen I and the CD-Bonn potentials, respectively.
The solid and dash-dotted curves are $V_M$ ($\Lambda$ = 200 MeV),
the solid one results from the Nijmegen I and the CD-Bonn potentials,
while the dash-dotted from the Reid93.
}
 \end{minipage}
 \hspace{6mm}
 \begin{minipage}[t]{75mm}
 \includegraphics[width=75mm]{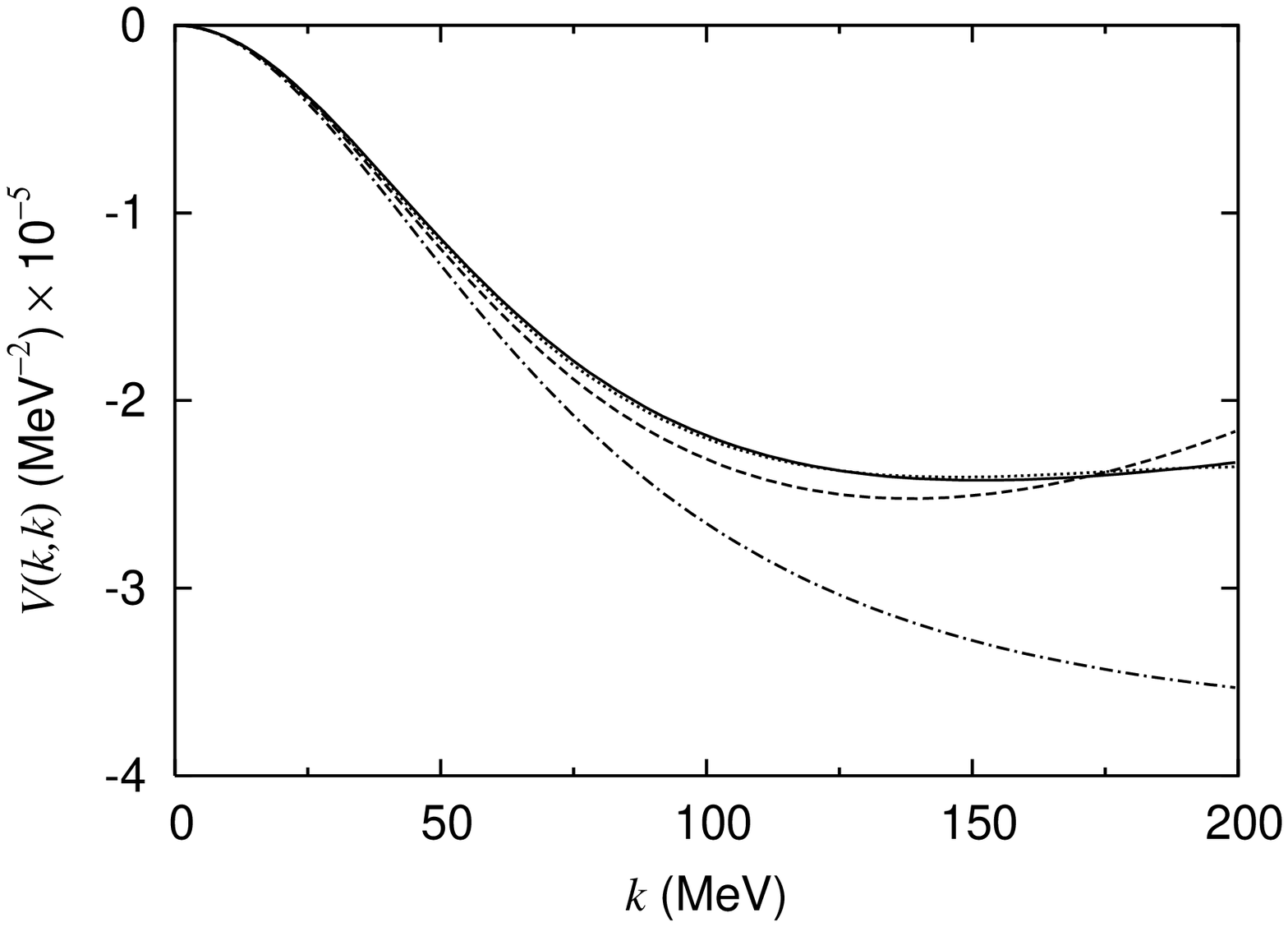}% Here is how to import EPS art
 \caption{\label{fig2}
$NN$-potential
$\bra{\td1}V\ket{\ts1}$.
The solid curve is
$V_M$($\Lambda$=200MeV)
resulting from the CD-Bonn potential.
The dashed (dotted) curve is a simulation of $V_M$ using the OPEP plus
one (two) contact interaction(s). The dash-dotted curve is the OPEP.
 }
 \end{minipage}
\end{figure}

In Fig.~\ref{fig1}, shown are diagonal components of 
$\bra{\1s0}V\ket{\1s0}$ in the momentum space. 
There is the clear model dependence among \vph, which almost
disappears after the model-space reduction down to $\Lambda$ = 200 MeV.
The obtained $V_M$ is accurately simulated by the OPEP and a few
contact interactions, which we do not show here. 
In Fig.~\ref{fig2}, diagonal components of
$\bra{\td1}V_M\ket{\ts1}$ ($\Lambda$=200MeV)
obtained from the CD-Bonn potential is simulated using
the OPEP and one or two contact interactions. We see that the
simulation is very accurate.
\veft\ obtained in this way also accurately simulates the
off-diagonal components of $V_M$.

\section{Discussion and summary}
We have seen in the previous section that the scenario for the
relation between \veft\ and \vph\ is realized. 
Although we have shown the relation employing a rather small model space,
if the TPEP and higher order OPEP are properly included in \veft, then
the relation should hold true for a larger model space as
well; probably safe up to $\Lambda\sim$ 400 MeV.
In a certain large model space where the very details of the TPEP is
considerable, {\it i.e.}, a phenomenological model cannot mimic the TPEP
any more, the relation may not hold true.
We have been concerned with only the strong nucleon-nucleon
interaction. However,
this kind of relation probably exists also in other nuclear operators,
such as electroweak currents and pion production operators.

Now, based on our result, we discuss in the following some features of
NEFT which the traditional approach does not have; the discussion is not
always restricted to the nuclear force.
One is that nuclear operators based on NEFT are constructed
perturbatively.
Therefore, one can systematically improve the accuracy, and can discuss
theoretical uncertainty. 
We note that, for a convergent perturbation, it is necessary to define
an operator in a suitably small model space.
We also note that the convergence of
the NEFT-based perturbative expansion is not always good for some
processes.

Another is that NEFT-based operators are model independent,
which has been often claimed.
However, the claim is based on the procedure of deriving the nuclear
operators, which is not a quantitative argument.
In this work, we quantitatively showed that \veft\ does not have the
dependence on modelling the details of short-distance physics.
This is a consequence of viewing the system roughly, thereby ignoring
the detailed structure of the short-distance physics.
It is noted, therefore, a description of a system based on NEFT
is not always better than those based on phenomenological models.
In many cases, they give essentially the same low-energy observables
because they are
equivalent through the RG. NEFT may have superiority in cases
where detailed information about multi-pion exchange mechanism play an
important role.

Finally, with NEFT, one can describe a system more efficiently
and simply, compared to the phenomenological models.
The efficiency is due to the perturbative calculation, and the
simplicity is due to the ignorance of the details of the short-distance
physics.

To summarize, we considered the nuclear force from the viewpoint of the
RG, and showed that there exists the equivalence relation between \veft\
and \vph. We simultaneously showed that \veft\ does not have the model
dependence due to the details of the short-distance physics, and that
the simple contact interactions in \veft\ accurately describe the
short-distance physics in a certain small model space. 
Based on the result, we discussed the new features in NEFT.


\begin{thebibliography}{9}
\bibitem{weinberg}
S. Weinberg, Phys. Lett. B251 (1990) 288.

\bibitem{mine}
S. X. Nakamura, Prog. Theor. Phys. 114 (2005) 77.

\bibitem{birse}
M. C. Birse, J. A. McGovern and K. G. Richardson, 
Phys. Lett. B464, (1999) 169.
\end{thebibliography}
\end{document}